\documentclass[10pt]{article}
\usepackage{emulateapj}

\begin{document}

\title{Photospheric Metals in the {\it FUSE} Spectrum\\
of the Subdwarf B Star PG0749+658}

\author{R. G. Ohl,\altaffilmark{1,2}
P. Chayer,\altaffilmark{1,3}
and H. W. Moos\altaffilmark{1}}
\altaffiltext{1}{Department of Physics and Astronomy, 
      The Johns Hopkins University, Baltimore MD 21218}
\altaffiltext{2}{Astronomy Department,
University of Virginia, P.O. Box 3818, Charlottesville, VA 22903}
\altaffiltext{3}{Primary affiliation: Department of Physics and Astronomy,
University of Victoria, P.O. Box 3055, Victoria, BC, V8W 3P6, Canada}

\begin{abstract}

We present an abundance analysis of 
the far-ultraviolet (905--1187~\AA) spectrum of
the subdwarf B star PG0749+658, obtained by 
the {\it Far Ultraviolet Spectroscopic Explorer (FUSE)}.  The data have 
a resolution of about 
${\rm R} = \lambda / \Delta \lambda = $ 12000--15000 (20--25 ${\rm km/s}$).  
We determine C, N, Si, P, S, Cr, Mn, Fe, Co, 
and Ni abundances, and upper limits
on the abundance of Cl and V, using a grid 
of synthetic spectra based on a LTE stellar atmosphere model.  He, C, N, Si, 
and Cl are depleted by a factor of $\gtrsim$10 with respect to solar, while
P, S, and Fe are diminished by less than a factor of 10.  We measure a solar
abundance of Cr, Mn, and Co and a Ni enhancement of $\sim$0.6~dex.  We compare 
these values to predictions based on 
radiative levitation theory.  
The radial velocity of the foreground interstellar
material has one dominant component, and coincides with that of the 
photospheric lines.

\end{abstract}


\section{INTRODUCTION}     \label{sec: sdintro}

PG0749+658 was discovered by the Palomar-Green survey as an 
ultraviolet-excess object (Green, Schmidt, \& Liebert~1986) and classified
as a subdwarf B star (sdB).  In their study of sdB candidate stars found
in the Palomar-Green survey, Saffer et al.~(1994) measured the star's 
effective temperature, gravity, and He abundance (Table~1).
With an effective temperature of 24,600~K, PG0749+658 is near the cool end
of the sdB population.

SdBs lie on the extreme horizontal branch (EHB) of the HR diagram and are
progenitors of low-mass white dwarfs (Dorman, Rood, \& O'Connell~1993).  They
are core He-burning stars with high effective temperatures
(20,000~K $\leq T_{\rm eff} \leq$ 40,000~K) and gravities 
($\log g \simeq$~5--6), and small H-rich envelopes 
($M_{\rm env} < 0.05 M_{\odot}$) (see, e.g., Saffer et al.~1994 
and references therein).  Precisely how EHB stars evolve to their location on 
the HR diagram 
is controversial.  Proposed binary scenarios draw on mass loss 
(Mengel, Norris, \& Gross~1976) or mergers (Iben 1990).  Single star theories 
involve heavy mass loss on 
the red giant branch (RGB), leaving the star on the zero age HB with
a low $M_{\rm env}$ and high $T_{\rm eff}$ (see, 
Yi, Demarque, \& Oemler~1997 and references therein).  
This interpretation invokes mass loss on the RGB that is a sensitive 
function of metallicity.

Although the evolutionary path followed by the progenitors of sdB stars is 
unclear, sdBs are important to understanding the late stages of stellar
evolution, with implications for the evolution of galaxies.  
Early vacuum ultraviolet observations of E/S0 galaxies and the bulges of 
spiral galaxies revealed that these old populations, which were thought to
harbor no significant number of stars hotter than their main-sequence 
turn-offs ($T_{\rm eff} \simeq$~5,500~K), are copious UV emitters 
(see O'Connell~1999 for a review).  
For the leading theories which address the ``UV excess'' 
(e.g., Yi, Demarque, \& Oemler~1997), sdB stars represent 
only one candidate for the source of the emission.  
However, sdBs have lifetimes much 
longer than other EHB stars, and dominate the UV excess 
(Dorman, Rood, \& O'Connell~1993; Brown et al.~1997).

Some sdB stars, named for the prototype EC14026, are observed to pulsate 
(Kilkenny et al.~1997).  At present, 14 sdB pulsators have been discovered
and all show multi-periodic, high-frequency luminosity variations with typical
periods of $\sim 100$--300~sec and amplitudes of $<1$--25\% (see, e.g., 
Koen et al.~1998, O'Donoghue et al.~1999, and 
Bill\`{e}res et al.~2000).  The
peculiar iron abundance profile, produced in these high gravity stars by
diffusion processes, is thought to play an important role in the
pulsation mechanism (Charpinet et al.~1996 and~1997).  Although PG0749+658 is 
not expected to pulsate because its effective
temperature is somewhat low (G. Fontaine~2000, private communication), the
study of its FUV spectrum provides an opportunity to test the diffusion
theory, which is one of the major ingredients in explaining the pulsation
of sdB stars.  Understanding the pulsation mechanism can
potentially be used to shed light on the interior of sdB stars and, thus,
their evolutionary status.

Abundance anomalies are observed in the atmospheres of sdB stars.  Low
resolution optical spectroscopy reveals that the abundance of helium in sdB
stars is typically a factor of ten smaller than the solar abundance (see, 
e.g., Moehler, de~Boer, \& Heber~1990).  On
the other hand, high resolution optical and UV spectra show that elements
heavier than He may or may not be underabundant.  For
instance, C and Si are underabundant in sdB stars, while
the N abundance is nearly solar over the effective temperature range
covered by sdB stars (Lamontagne, Wesemael, Fontaine, \& Sion~1985; 
Heber~1991).  The processes of gravitational settling, radiative
levitation, and mass loss are thought to be responsible for the
observed abundance anomalies in these EHB stars, which have gravities
larger than those of 
main sequence stars (Michaud et al.~1985; Bergeron et al.~1988; 
Fontaine \& Chayer~1997; Unglaub \& Bues~1998).

We present the far-ultraviolet (FUV; 905--1187~\AA) spectrum of 
PG0749+658 obtained by the {\it Far Ultraviolet
Spectroscopic Explorer (FUSE)}.  This is the first high-resolution spectrum 
of a sdB shortward of Ly$\alpha$.  
In this paper we perform an abundance analysis and we compare our results
with radiative levitation theory.  Furthermore, this study is a 
test case in support of future {\it FUSE} subdwarf observations.

\section{OBSERVATIONS AND DATA REDUCTION}    \label{sec: sdfuvobs}

The {\it FUSE} mission and instrument performance are covered by 
Moos et al.~(2000) and Sahnow et al.~(2000), respectively.

This spectrum was obtained as a part of In-Orbit Checkout program
I904 on 9~November~1999 using the high-throughput
apertures ($30 \arcsec \times 30$ \arcsec; LWRS) on all four spectrograph 
channels and recorded in ``histogram'' mode for a total exposure time of
17,616~sec.  On 7~January~2000, six additional exposures were obtained 
in ``time tag'' mode for a total of 9,409~sec, but only data from the
LiF1 and LiF2 channels were usable.  
The spectra were processed by the {\it FUSE} pipeline (version number 
1.6.8).  No flat-fielding was performed, and the dominant
statistical uncertainty is the fixed pattern noise associated with the
detectors.  Astigmatism and 
dead-time corrections were not applied, but rough photometric and wavelength
calibrations were performed.  For this analysis, we co-added individual 
exposures to increase the signal-to-noise ratio and corrected 
the spectra for the star's radial velocity 
(Table~1).  
We detect one radial velocity component of the foreground interstellar medium
and many interstellar ${\rm H}_{2}$ and metal lines are present.

Selection of the {\it FUSE} channel used for fitting a given photospheric line 
was based 
on the local data quality (i.e., the signal-to-noise ratio and the 
presence of artifacts).  
A wavelength solution has been applied to the data, but it is inaccurate on
small scales ($\lesssim 0.1$~\AA).  
Before fitting each stellar line, we created a local wavelength solution over 
$\sim 6$--10 \AA\ by linearly interpolating between
the known locations of interstellar features and the 
photospheric lines of interest.  
These local corrections to the wavelength scale
greatly improved the fit used to determine abundance 
(Section~\ref{sec: sdresults}).

The uncertainty in the absolute wavelength calibration dominates the
error in the 
measurement of radial velocity.  However, measurements of the
{\it relative} locations of photospheric and interstellar lines indicate that
the radial velocity of PG0749+658 and the dominant component of the foreground 
interstellar medium are essentially 
equal, within the uncertainty set by local distortions in the wavelength
scale ($<$~30~${\rm km/s}$).  This makes the interstellar 
lines useful for correcting the wavelength scale, but it is harder to 
disentangle blends of interstellar and photospheric lines.

\section{RESULTS}     \label{sec: sdresults}

We computed a stellar model under the assumption of
local thermodynamic equilibrium (LTE), and a pure H and He composition using the
stellar atmosphere code 
TLUSTY (Hubeny \& Lanz~1995) and parameters derived by 
Saffer et al.~(1994) (Table~1).  
Using this model, we generated a 
grid of synthetic spectra for each photospheric line using the SYNSPEC code
(I.~Hubeny) and atomic data provided by R.~L.~Kurucz.  
The grid covered a range in metal abundance and local instrument 
resolution.  We then fit each line individually using a chi-square 
minimization technique, where, for a given wavelength range, 
the abundance, resolution, and normalization were optimized.  
To fit each line, we modeled a $\sim 6$--10~\AA\ range in the spectrum 
centered on the feature of interest.  We chose spectral features 
based on the strength of the transition, the
local data quality, and the presence of ``contaminating'' interstellar 
absorption lines.  We then checked that the derived abundance reproduced other
lines of the element observed elsewhere in the spectrum.
The abundances by number with respect to H are listed 
in Table~2 for all of the photospheric lines detected in the {\it FUSE} 
spectrum.  
A model spectrum using all of our determined abundances is shown in 
Figure~1.

The formal, $1 \sigma$ 
uncertainties associated with the output of the fitting process are much
lower than the systematic errors:  The uncertainties in the flux and 
wavelength calibrations, errors in the oscillator strength used to 
calculate our synthetic spectra, uncertainty in the atmospheric parameters
(Table~1), and the potential blending of the photospheric 
lines with interstellar absorption 
features represent significant sources of systematic error.  
Furthermore, we neglect non-LTE 
effects in the calculation of our model atmosphere, which may also 
contribute to this error.  Studies examining departure from LTE indicate
that the impact on our model line profiles is probably not critical to
this abundance analysis for this cool sdB (Baschek et al.~1982; 
Heber et al.~1984).  
We neglect microturbulence in our abundance analysis.  
We estimate the combined systematic errors to impact our abundance 
measurements on about the $\pm$0.3~dex-level.

The photospheric abundances and the predictions of radiative levitation
theory are listed in Table~2.  
The predicted equilibrium abundance for a given element is computed within
the framework of the radiative levitation theory developed by 
Chayer et al.~(1995a, 1995b).  The computations are performed at the 
Rosseland optical depth, $\tau_{\rm ross} = 2/3$, for a hydrogen-rich
stellar atmosphere model with $T_{\rm eff} = $ 24,600~K and
$\log g = 5.5$.  The equilibrium 
abundance was calculated only for the elements available in the TOPBASE
data bank (i.e., those with cosmic abundances by number greater than 
$\sim 10^{-6}$ relative to H; 
Cunto \& Mendoza~1992).  
These theoretical values are also plotted for comparison to the 
observed PG0749+658 photospheric and solar 
abundances in Figure~2.  
The solar abundances are from Anders \& Grevesse~(1989).

\section{DISCUSSION}     \label{sec: sddiscuss}

The FUV spectral analysis of PG0748+658 is summarized in 
Figure~2, where the observed, predicted and solar
abundances are compared.  All of the elements are depleted with respect to 
solar, except 
Cr, Mn, Co, and Ni.  Among those that are depleted, He, C, N, Si, and Cl are 
diminished by a factor of $\gtrsim$10, while P, S, and Fe are 
depleted by less than a factor of 10.  We observe an enhancement of Ni
of $\sim$0.6~dex.  We do 
not measure over-abundances of P and Cl, as reported by 
Baschek \& Sargent~(1976) and Heber~(1991), respectively, in their 
studies of cooler HB stars of spectral type B (HBB).

The measured abundances are generally consistent with predictions based on
radiative levitation theory, except for He and Si.  
The under-abundance of He is not explained by this theory 
alone.  On the other hand, 
Fontaine \& Chayer~(1997) and Unglaub \& Bues~(1998) suggest that an 
outward velocity field, created by a weak stellar wind, may support
significant amounts of He in the photosphere of sdB stars.

The observed Si abundance falls below that expected from theory by about
two orders of magnitude.  This result is consistent with that of 
Bergeron et al.~(1988), 
who show that radiative levitation theory does not predict the observed 
under-abundance of Si in sdB stars.  Calculations by Michaud et al.~(1985) 
indicate that this anomaly may be explained by the presence of a weak 
stellar wind.

The predicted and observed Fe abundance agree well, which 
supports the application of radiative levitation theory toward 
predicting the amount of Fe in the atmospheres of hot subdwarfs.

Lamontagne et al.~(1987) observed PG0342+026 with {\it IUE} and 
determined C, N, 
and Si abundances.  This star and PG0749+658 have very similar atmospheric
parameters.  Averaging the parameters obtained by 
Lamontagne et al.~(1987), Moehler, de Boer, \& Heber~(1990), 
and Saffer et al.~(1994) gives 
$T_{\rm eff} = 24$,800~K, $\log g = 5.3$, and $\log {\rm He/H} = -2.6$ 
for PG0342+026.  The 
C and Si abundances of the two stars are in good agreement.  
However, the N abundance 
is about 1.7~dex higher for PG0342+026, but the uncertainties in the 
measurements for both stars are large.


\acknowledgments

We gratefully acknowledge the many contributions of the {\it FUSE} 
Instrument and
Science Operations Team.  We thank I.~Hubeny for providing his stellar
atmosphere and spectral synthesis codes, and R.~L.~Kurucz for providing
his atomic transition data.  PC is a Canadian representative to the {\it FUSE}
Project supported by CSA under PWGSC.  This work was supported by NASA
grant NAS5-32985.


\begin{deluxetable}{lcc}
\tablecolumns{8}
\tablewidth{0pt}
\tablecaption{Target Summary for PG0749+658}
\tablehead{
\colhead{Quantity} &
\colhead{Value} &
\colhead{ref}
}
\startdata
 Spectral Type & sdB & 1 \\
 $V$ & 12.121 & 2 \\
$B-V$ & $-$0.106 & 2 \\
$V-R$ & 0.021 & 2 \\
$R-I$ & 0.072 & 2 \\
$T_{\rm eff}$ & 24,600 K & 3 \\
$\log g$ & 5.5 & 3 \\
$\log {\rm He / H}$ & $-$2.4 & 3 \\
$V_{\rm rad}$ & $-27.3 \pm 6.9$ & 4 \\
\enddata
\tablerefs{(1) Green, Schmidt, \& Liebert 1986; (2) Allard et al. 1994;
(3) Saffer et al. 1994; (4) Saffer, Livio, \& Yungelson 1998.}
\end{deluxetable}

\clearpage

\begin{deluxetable}{lcc}
\tablecolumns{8}
\tablewidth{0pt}
\tablecaption{Ionic Abundance Analysis}
\tablehead{
\colhead{Ion} &
\colhead{$\epsilon$\tablenotemark{a}} &
\colhead{$\epsilon$\tablenotemark{a}} \\
\colhead{ } &
\colhead{(observed)} &
\colhead{(predicted)} 
}
\startdata
C II & 6.59 & 6.78 \\
C III & 7.08 & 6.78\\
N III & 6.93 & 6.20\\
Si III & 5.16 & 7.30\\
Si IV & 5.46 & 7.30\\
P III & 4.69 & $\cdots$ \\
S II & 6.00 & 6.23 \\
S III & 6.76 & 6.23\\
S IV & 6.55 & 6.23\\
Cl III & $< 3.5$ & $\cdots$ \\
V III & $< 4.0$ & $\cdots$ \\
Cr III & 5.37 & $\cdots$ \\
Mn III & 5.29 & $\cdots$ \\
Fe III & 7.32 & 7.26\\
Co III & 4.65 & $\cdots$ \\ 
Ni III & 6.88 & $\cdots$ \\ 
\tablenotetext{a}{$\epsilon = 12 + \log [ {\rm N(A)/N(H)} ]$.}
\enddata
\end{deluxetable}
 
\clearpage


\clearpage

\figcaption{{\it FUSE} spectrum (solid line) 1120--1130~\AA\ 
and model using all of 
the observed metal abundances given in Table~2 (dotted line).  Lines with
equivalent widths larger than 20~m\AA\ are identified above the 
spectrum.  The vertical lines below the spectrum mark the location of
interstellar Fe~II absorption features.}

\figcaption{Comparison of the observed and expected abundances of elements 
found in the photosphere of the sdB star PG0749+658.  
Systematic errors for both the observed and
predicted abundances are about $\pm 0.3$~dex.  
The data place stringent upper limits on the abundances of Cl and V (arrows).  
The predicted abundances were computed for elements available in the TOPBASE
data bank, i.e., He, C, N, Si, S, and Fe.  Where multiple species are 
available for a given element, the average of the derived abundances
is shown.  The He abundance is from
Saffer et al.~(1994).}


\newpage

\clearpage

\addtocounter{figure}{-2}

\begin{figure}[t]
\includegraphics{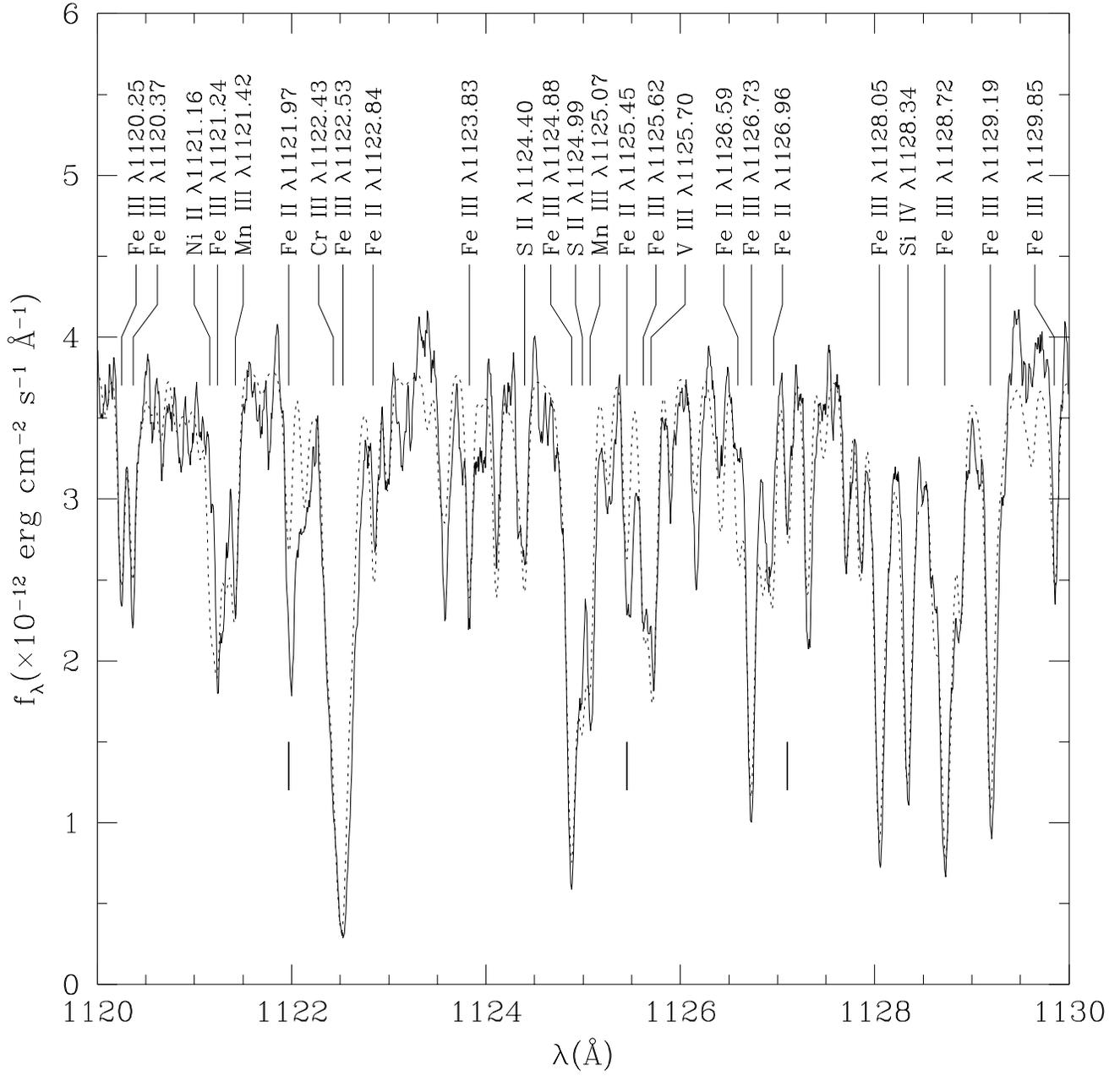}
\vspace{15.0cm}
\caption{{\it FUSE} spectrum (solid line) 1120--1130~\AA\ 
and model using all of 
the observed metal abundances given in Table~2 (dotted line).  Lines with
equivalent widths larger than 20~m\AA\ are identified above the 
spectrum.  The vertical lines below the spectrum mark the location of
interstellar Fe~II absorption features.}
\label{fig: metals}
\end{figure}

\newpage

\clearpage

\begin{figure}[t]
\includegraphics{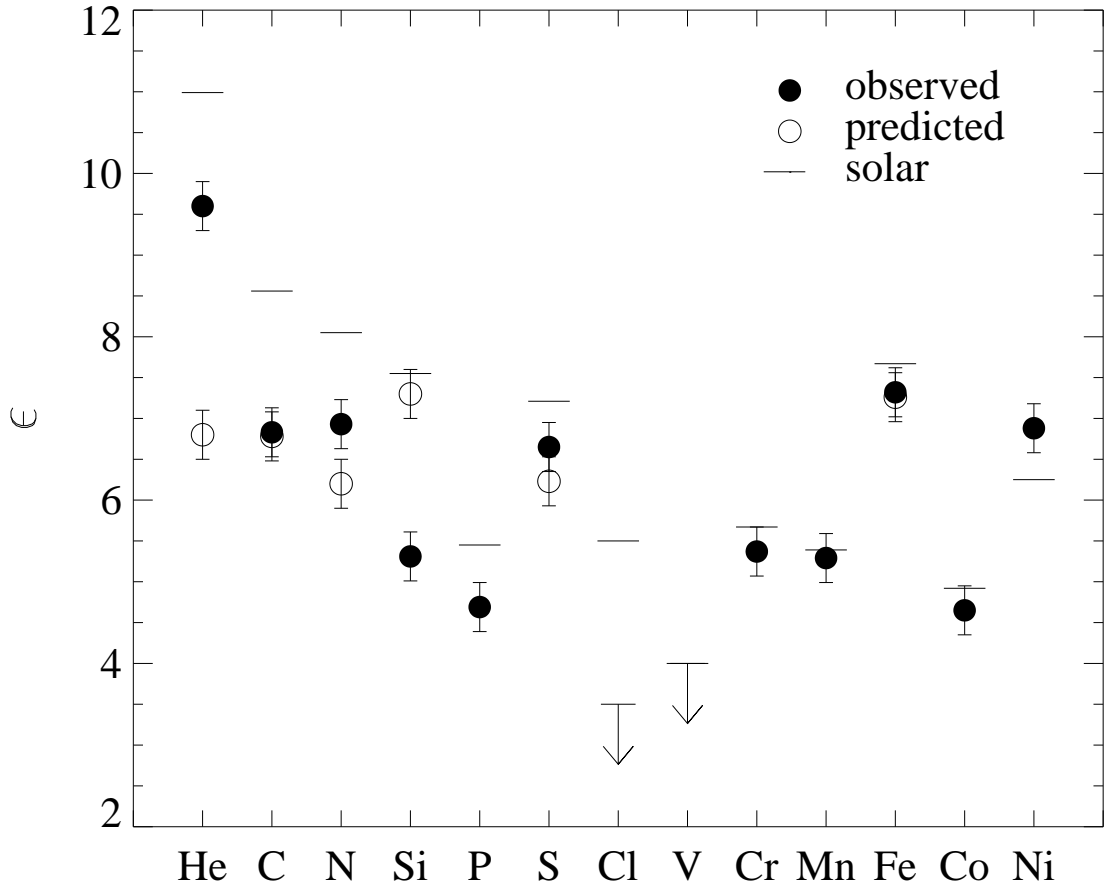}
\vspace{15.0cm}
\caption{Comparison of the observed and expected abundances of elements 
found in the photosphere of the sdB star PG0749+658.  
Systematic errors for both the observed and
predicted abundances are about $\pm 0.3$~dex.  
The data place stringent upper limits on the abundances of Cl and V (arrows).  
The predicted abundances were computed for elements available in the TOPBASE
data bank, i.e., He, C, N, Si, S, and Fe.  Where multiple species are 
available for a given element, the average of the derived abundances
is shown.  The He abundance is from
Saffer et al.~(1994).}
\label{fig: abund}
\end{figure}

\end{document}